\pdfoutput=1
\documentclass[doublecol,a4paper]{epl2} 

\usepackage{amsmath,amssymb,graphicx}
\usepackage{times}
\newcommand{\fig}[2]{\includegraphics[width=#1]{./figures/#2}}
\newcommand{\Fig}[1]{\includegraphics[width=\columnwidth]{./figures/#1}}
\newlength{\bilderlength}

\newcommand{\rme}{{\mathrm{e}}}
\newcommand{\rmd}{{\mathrm{d}}}

\newcommand{\DRAFT}[1]{}

\begin{document}
\bibliographystyle{KAY}
\title{
Height fluctuations of a
contact line: a direct measurement of the renormalized disorder correlator}
\author{Pierre Le Doussal and Kay
J\"org Wiese\\
\mbox{\rm\it CNRS-Laboratoire de Physique
Th{\'e}orique de l'Ecole Normale Sup{\'e}rieure, 24 rue Lhomond, 75231
Paris Cedex, France}\medskip \\
 Sebastien Moulinet \and Etienne Rolley\\
  \rm \it Laboratoire de Physique Statistique, (ENS,CNRS,  Universit\'{e}s Paris 6 et Paris 7), 24
rue Lhomond, 75005 Paris, France.\vspace{-8mm}}
\institute{}
\shortauthor{Pierre Le Doussal, Kay J\"org Wiese, Sebastien Moulinet, Etienne Rolley}
\date{\small\today}
\abstract{
We have measured the center-of-mass fluctuations of the height of a contact
line at depinning for two different systems: liquid hydrogen on a rough cesium substrate and isopropanol on
a silicon wafer grafted with silanized patches. The contact line is subject to a confining quadratic well,
provided by gravity. From the second cumulant of the height fluctuations, we measure the
renormalized disorder correlator $\Delta(u)$, predicted by the
Functional RG  theory to attain a fixed point, as soon as the
capillary length is large compared to the Larkin length set by the
microscopic disorder. The experiments are consistent with the asymptotic form for
$\Delta(u)$ predicted by Functional RG, including a linear cusp at $u=0$. The observed small deviations could be used as a probe of the underlying physical processes. 
The third moment,
as well as avalanche-size distributions are measured and compared to
predictions from Functional RG.}
\maketitle
\enlargethispage{2mm}

A direct measurement of the fixed-point function $\Delta(u)$, the so-called
renormalized disorder correlator, which
plays a central role in the Functional RG theory (FRG) of pinned elastic systems, was
recently proposed \cite{LeDoussal} and verified in an exact
numerical determination of ground states for interfaces in various
types of disorders  \cite{MiddletonLeDoussalWiese2006}. The main idea
is to put the elastic system in a quadratic
potential well, which acts as a large-scale cutoff and makes the problem well-defined. The shift between the center of mass 
and the center of the
well
is proportional to the renormalized force and
its fluctuations are the
quantity computed in the FRG \cite{LeDoussal}. The results of \cite{MiddletonLeDoussalWiese2006} show a
remarquable agreement in the statics between the measured $\Delta(u)$
and the 1- and 2-loop predictions from the Functional RG
\cite{DSFisher1986,FRGdepinning,ChauveLeDoussalWiese2000a}. These ideas and numerical tests
have been extended to the depinning transition
\cite{LeDoussalWiese2006a,RossoLeDoussalWiese2006a}
in the case of local elasticity, and to reaction diffusion models \cite{BonachelaAlavaMunoz2008}.
Finally, a first-principle calculation of the
distribution of avalanches from the FRG was performed
and verified by numerics \cite{LeDoussalMiddletonWiese2008,LeDoussalWiese2008c,RossoLeDoussalWiese2009a}. An outstanding challenge is
to test these predictions in experiments.

\setcounter{topnumber}{1}
\setcounter{bottomnumber}{1}
\begin{figure}[b]
\centerline{\Fig{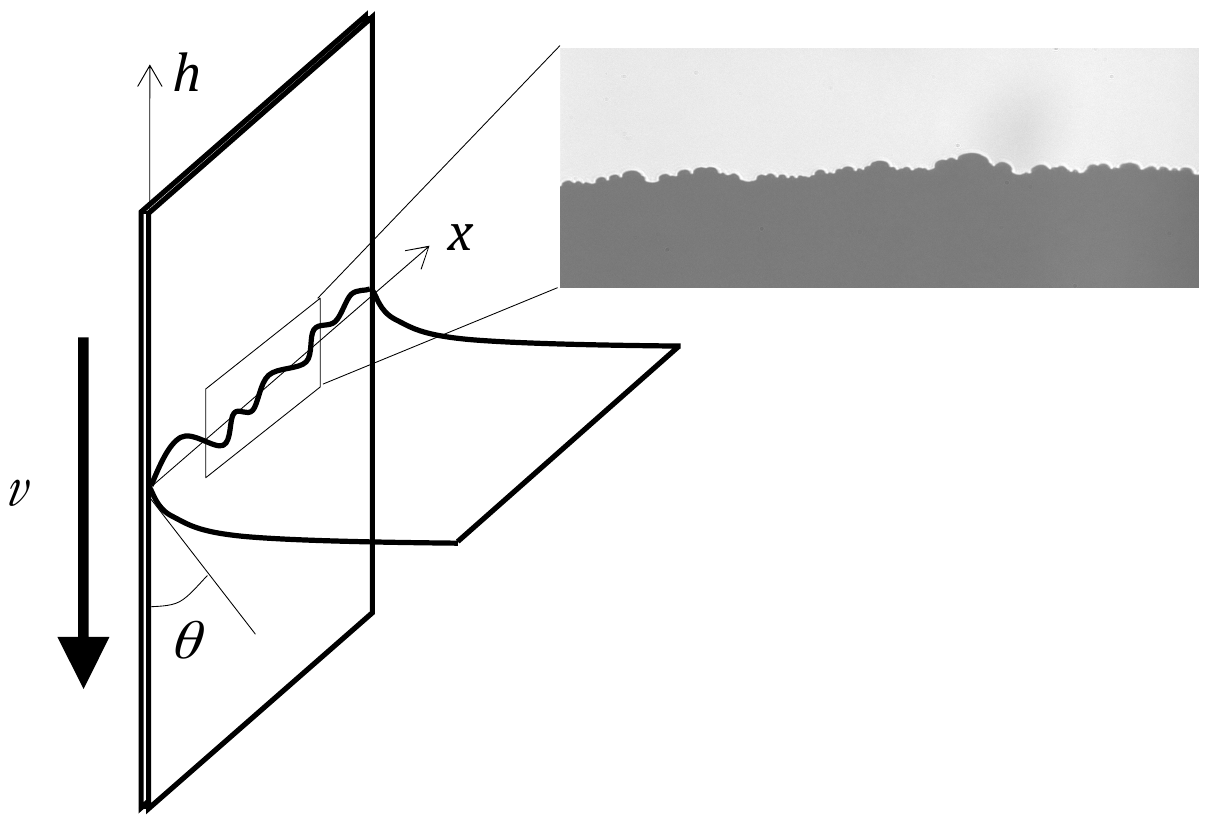}}
  \caption{Sketch of the experimental setup. The size of the image in the inset is 1.5 mm.
  }\label{setup}
\end{figure}

The depinning of the contact line of a fluid on a disordered substrate has been studied experimentally\cite{RolleyGuthmann2007,ZechFubelLeidererKlier2004,MoulinetGuthmannRolley2002,MoulinetRossoKrauthRolley2004}
and its critical scaling established. Since gravity naturally creates a quadratic well, this
raises the interesting possibility of measuring a FRG fixed-point
function for depinning or, conversely, to learn more about the
physical system using these (universal) fluctuations as a new probe. To do so, the capillary length, which
provides the well, does not need to be larger than the measurement
scale. In fact, the finite  capillary length is
used as an advantage.

Consider a fluid in a large reservoir and its contact line (CL) on a
plate, parametrized by $(x,u(x))$ within the plate (fig.~\ref{setup}). Its energy
can be modeled as
\begin{equation}\label{a1}
{\cal H}[u] = \int_{0}^{L} \rmd x\, \frac{m^2}{2} \left[u(x)-w\right]^2  + V (x,u(x)) + {\cal E}[u]\ ,
\end{equation}
where ${\cal E}[u]$
is a non-local elastic energy invariant under $u(x) \to
u(x) +\mbox{const.}$ The pinning force $-\partial_u V(x,u)$ is a local quenched random function.
$L$ is the total length of the contact line. 
\enlargethispage{2mm}
Additional boundary terms, not written here, affect the line near
the boundaries $x=0$ and $x=L$. The position  $w$ of the center of the well with respect to the plate, which is also the equilibrium position of the CL
without disorder, is fixed by the height in the reservoir far away from the contact line $w_{\infty}$, up to a constant shift $w = w_{\infty}+\sqrt{2} L_c \sqrt{1-\sin\theta}$, where $L_c=\sqrt{\gamma/ \rho g}$ is the capillary length and $\theta$ the contact angle. In the simplest model of the fluid surface 
 the elastic energy ${\cal E}[u]$ 
in Fourier space is
${\cal E}[u]=\frac{1}{2} \int \frac{\rmd k}{2 \pi} \epsilon_k u_{-k} u_k$ , with an elastic kernel
$\epsilon_k= m^2 \tilde \epsilon_\theta(k/k_\theta)$. The curvature of the quadratic well is
$m^2 = \gamma k_\theta$ with $k_\theta=\sqrt{2} \sin(\theta)/(L_c \sqrt{1+\sin \theta})$. The scaling function
$\epsilon_\theta(x)$ is often approximated by $\tilde \epsilon_{\pi/2}(x)=\sqrt{x^2+1}-1$, but  can be computed
for any $\theta$ \cite{LeDoussalWieseToBePublished}.

\begin{table}[t]
\begin{tabular}{cccccc}
\hline
\hline
            & $\theta$                &     $\gamma$      &        $L_C$      &         $\eta$           \\
            &    (degrees)              &     (mN$/$m)       &   (mm)            &       ($\times 10^{-3}$ Pa.s)                \\
\hline
$\mathrm{H}_2$    at 15K      & $\sim $ 40        &       2.79             & 1.94           &      0.021            \\
isobutanol & $\sim$ 40                &      20.9             &  1.64           &   4.0            \\
\hline
\hline
\end{tabular}
\caption{Wetting properties}
\label{tab}
\end{table}
\begin{figure}[t]
\centerline{\fig{6.5cm}{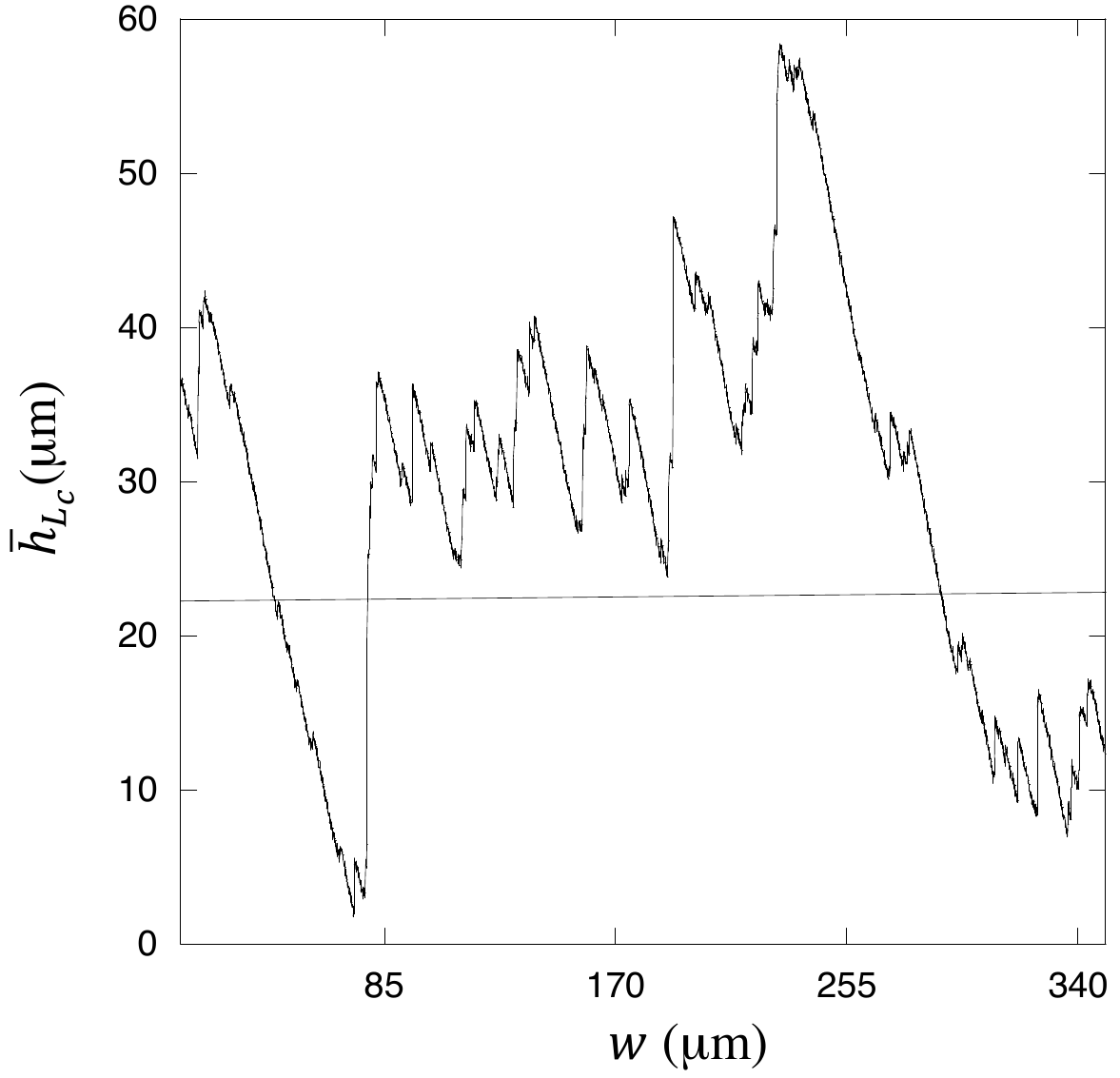}}
  \caption{Height of the contact line  $\bar{h}_{2L_C}(w)$ averaged over $2L_c$, as a function of the position $w$ of the plate (system: iso/Si). The fast depinning events (upwards) are clearly visible. Between them, the CL is advected downwards at the plate velocity $v$ (here 1 $\mathrm{ \mu m/s}$). The straight line is the reference level $h_0$.
  }\label{h(t)}
\end{figure}

One drives the system by slowly immersing the plate at a constant velocity $v$, hence $w=v t$. The center of mass
\begin{equation}
\bar u(t) := \frac{1}{L} \int_0^L \rmd x\, u(x,t)
\end{equation}
is fluctuating as a function of $w$, and contains valuable information about the system.
In particular, the second cumulant $\Delta(w)$, defined as
\begin{equation}\label{a3}
\Delta(v t):=L m^4 \langle  [\bar u(t+\tau) - v (t+\tau)][\bar u(\tau) - v \tau]  \rangle^c
\end{equation}
with $w=v t$
is exactly the renormalized disorder correlator computed by the FRG theory. Here and below $\langle \ldots \rangle$ denote translational averages in the direction of the motion of the CL.
Thus a direct measurement of $\Delta(w)$ is possible for the CL moving on a disordered substrate.

%

Two different systems have been used. The first one  is liquid hydrogen on a cesium substrate, denoted  $\mathrm{H}_2$/Cs hereafter. As in most previous experiments using $\mathrm{H}_2$/Cs, the Cs substrate is prepared at low temperature, yielding a rough surface with dense defects whose size is of the order of 10 nm \cite{RolleyGuthmann2007,ZechFubelLeidererKlier2004}. Such a small value leads to very small distortions of the CL and precludes any optical observation. We have thus annealed the Cs layer up to 250 K. This creates a large-scale structure (possibly corrugation) which is visible but cannot be characterized since its typical length scale is still below the optical resolution. Its typical length scale is in the micrometer range.

The second system is isobutanol on a silicon wafer, denoted iso/Si hereafter. We create a well-controlled disorder by photolithographic technics: The Si wafer is decorated by random silanized square patches ($10 \times 10\;\mu \mathrm{m}^2$) \cite{silane} which cover about $22 \% $ of the total area. As the silanized patches are less wettable than Si, each patch can pin the CL.

The relevant parameters (advancing contact angle $\theta$, liquid-vaper surface tension $\gamma$, capillary length $L_C$ and dynamic viscosity $\eta$) are listed in table \ref{tab} for both sytems. Note that contact angle and  capillary length are similar. In both systems defects are strong so that the Larkin length is set by the size $\xi$ of the defects; $\xi$ is macroscopic so that thermal activation is irrelevant \cite{RolleyGuthmann2007}.

A sketch of the experimental setup is shown in fig.~\ref{setup}. The  substrate is dipped into the liquid bath at constant velocity $v$. The CL is imaged with a standard progressive-scan CCD camera for velocities up to $10\;\mathrm{\mu m/s}$. For the room-temperature system, we have also used a high-speed camera in order to analyse the CL dynamics up to $v=1200 \;\mathrm{\mu m/s}$. For each time $t$, the CL profile is digitized to obtain the CL height  $h(x,t)$. The correspondence with the above notations is
\begin{equation}
h(x,t)-h_0 = u(x,t) - v t\ - \langle u(x,t) - v t \rangle \ .
\end{equation}
To define the reference level $h_0$ we first calculate the average height $\bar{h}_l(t)$
  over a CL length $l$: $\bar{h}_l(t) := \frac{1}{l} \int_0^l h(x,t)\, \rmd x$. An example 
is shown in fig.~\ref{h(t)}.
  Then $h_0$ is obtained as the time-averaged value of $\bar{h}_l(t)$.
  As the camera is fixed with respect to the liquid container, one expects $h_0$ to be constant.
  However, we allow for a slow drift of $h_0$ for two reasons: First,
as the volume of the container is finite, dipping the plate causes a
slight
 increase of the asymptotic level which results in a linear increase of the reference level $h_0$.
  Secondly, we find that the slow variations of $\bar{h}_l(t)$ are not linear,  due to large-scale variations of the plate properties (defect size, mean contact angle, etc.).
Hence, $h_0(t)$ is determined experimentally by fitting the whole
data set  $\overline{h}_l(t)$ by a polynomial function.
As we are interested in correlations over distances two
orders of magnitude smaller than the total swept distance, this only shifts the correlator $\Delta$ by
a constant.

In the following, instead of $h(x,t)$, we use ${\sf h}(x,w):=
h(x,w/v)-h_0$, where  $w= vt$, making it easier to compare runs at
different velocities $v$.
We define the experimentally  measured correlator $\hat \Delta$ as
\begin{equation} \label{defdelta}
\hat \Delta(w-w'):= \langle \overline{{\sf h}}_l(w)\overline{{\sf h}}_l(w')\rangle\ .
\end{equation}
\setcounter{topnumber}{2}
\begin{figure}[t]
\centerline{\fig{6cm}{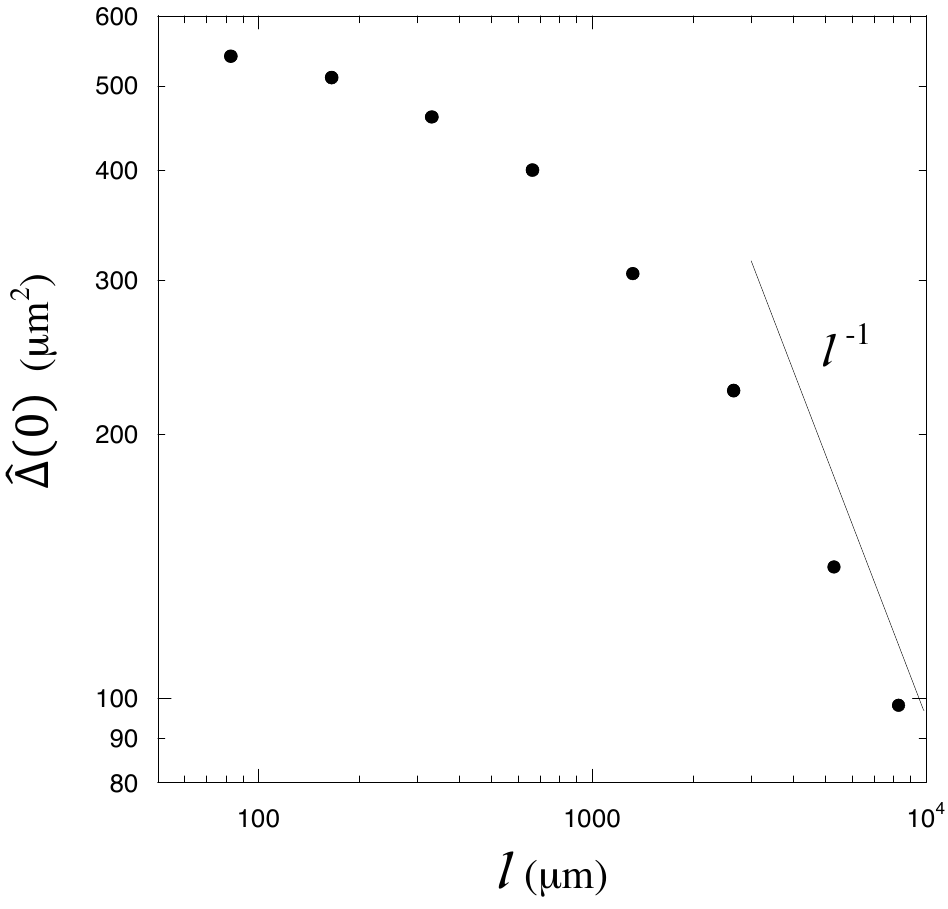}}
  \caption{Scaling of the correlator $\hat \Delta(0):= \langle \overline{\sf h}_l(w)\overline{\sf h}_l(w)\rangle$ as a function of the averaging length $l$ along the CL (here iso/Si). The asymptotic scaling  $l^{-1}$ is marginally reached for $l = L_c$.
  }\label{Delta(l)}
\end{figure}
\begin{figure}
\centerline{\fig{6.2cm}{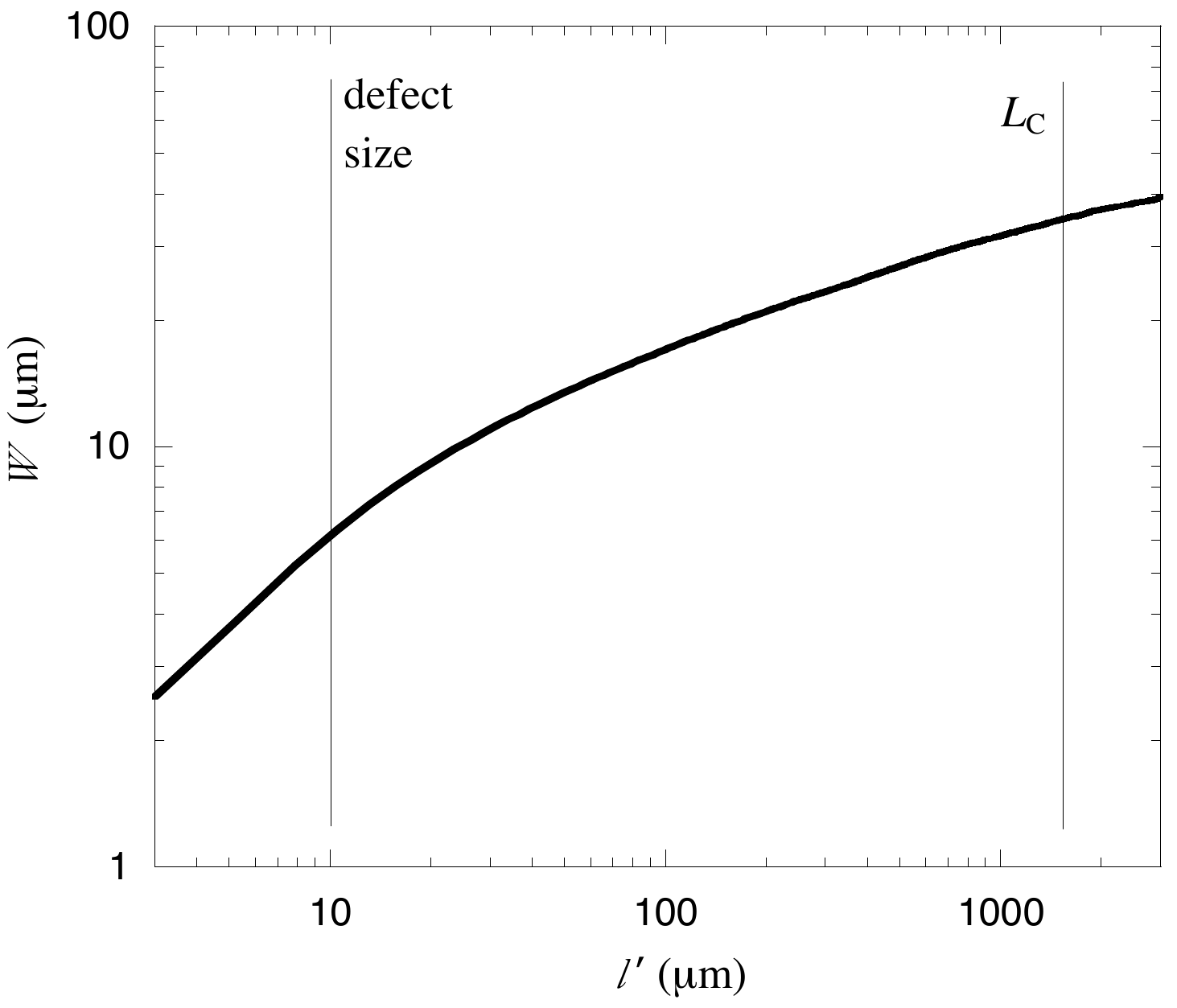}}
  \caption{Roughness $W(l')$ of the contact line for iso/Si. Due to the limited range between the defect size and the capillary length, no scaling is achieved.
  }\label{W(l)}
\end{figure}
\begin{figure}
\newlength{\inset}
\setlength{\inset}{6.5cm}
\Fig{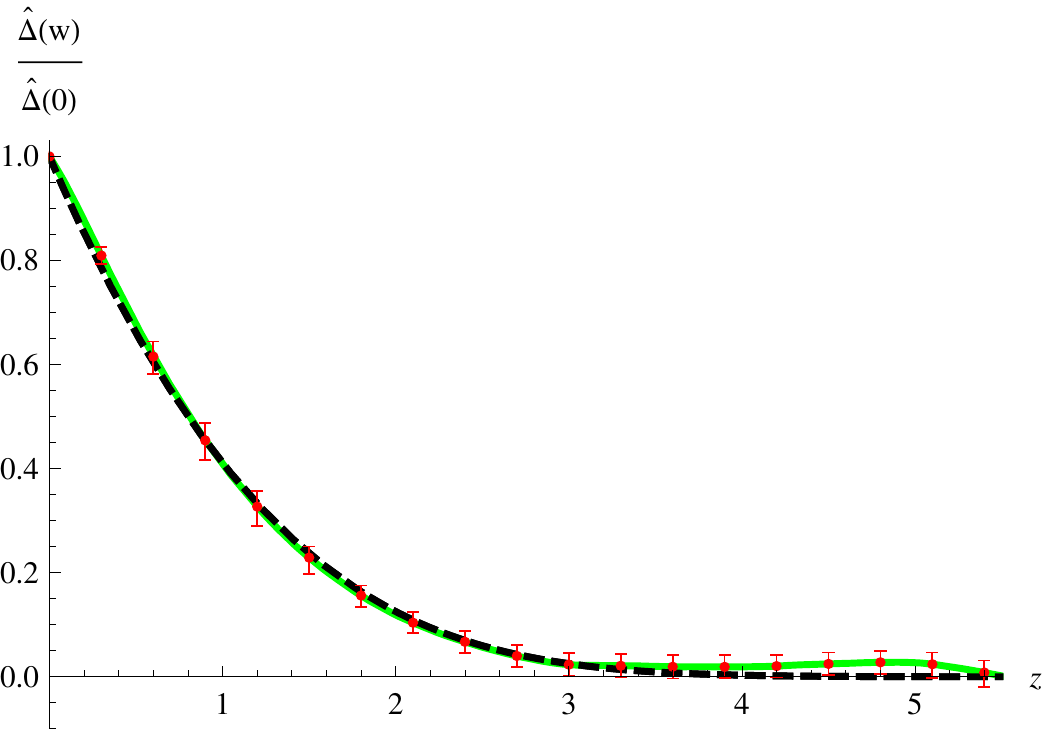}\hspace{-\inset}\raisebox{21.7mm}{\parbox[b]{\inset}{\fig{\inset}{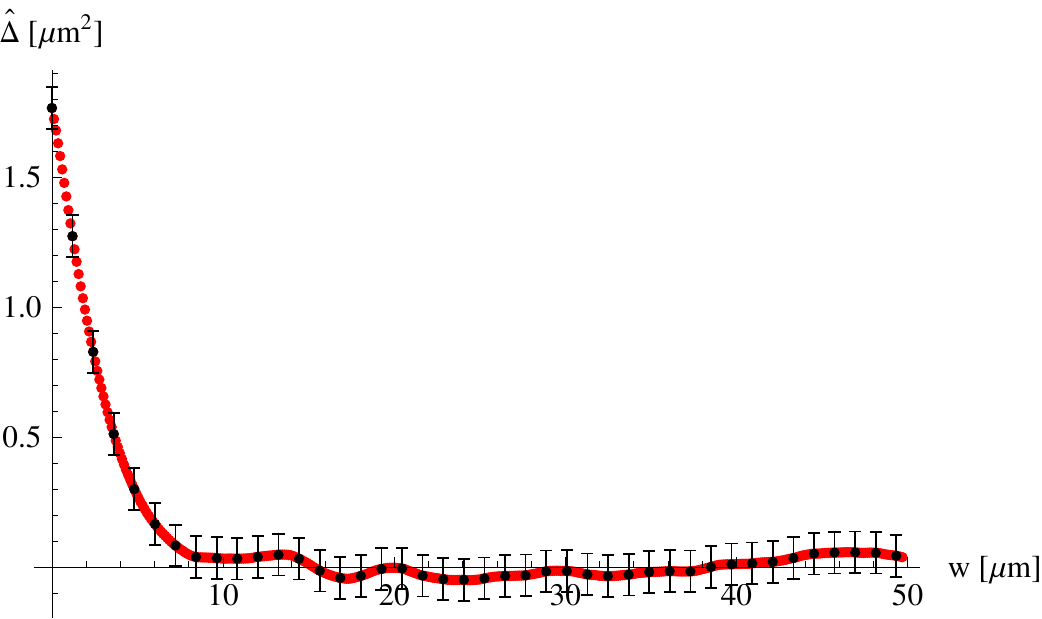}}}
\caption{Inset: The disorder correlator $\hat \Delta (w)$ for $\mathrm{H}_2$/Cs, with
error bars estimated from the experiment. Main plot:
The rescaled disorder correlator $\hat \Delta (w)/\hat \Delta (0)$ (green/solid) with error bars (red).
  The dashed line is the 1-loop result from equation
(\ref{Delta1loop}).}
\label{Delta:H2/Cs}
\setlength{\inset}{6.5cm}
\Fig{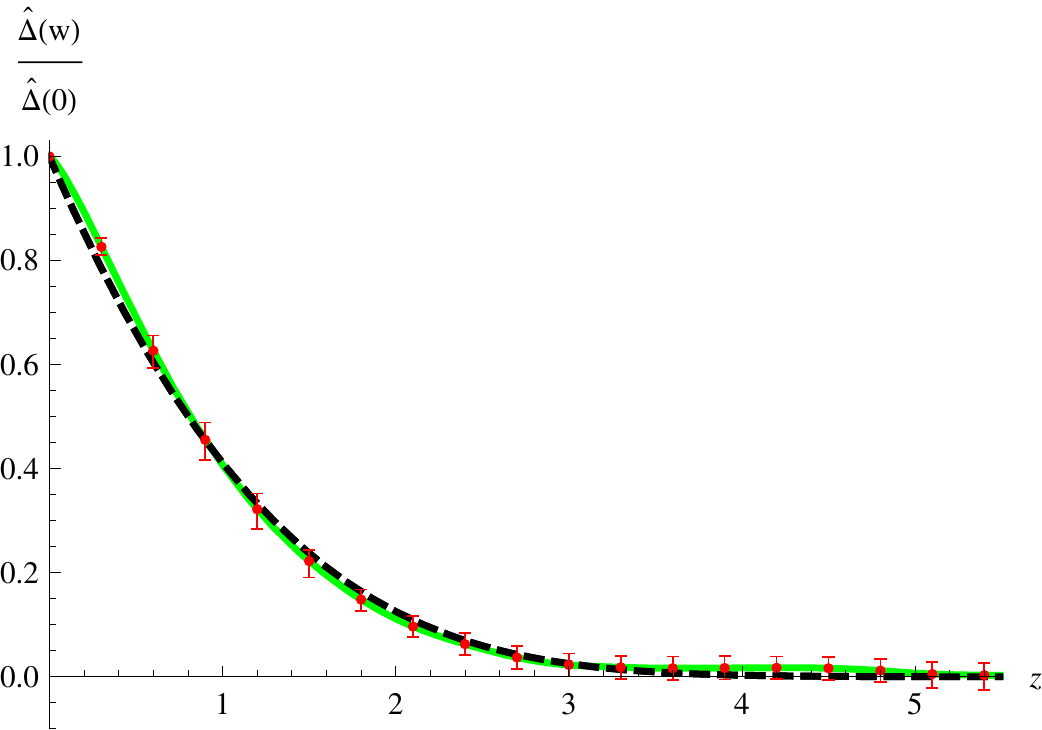}\hspace{-\inset}\raisebox{21.7mm}{\parbox[b]{\inset}{\fig{\inset}{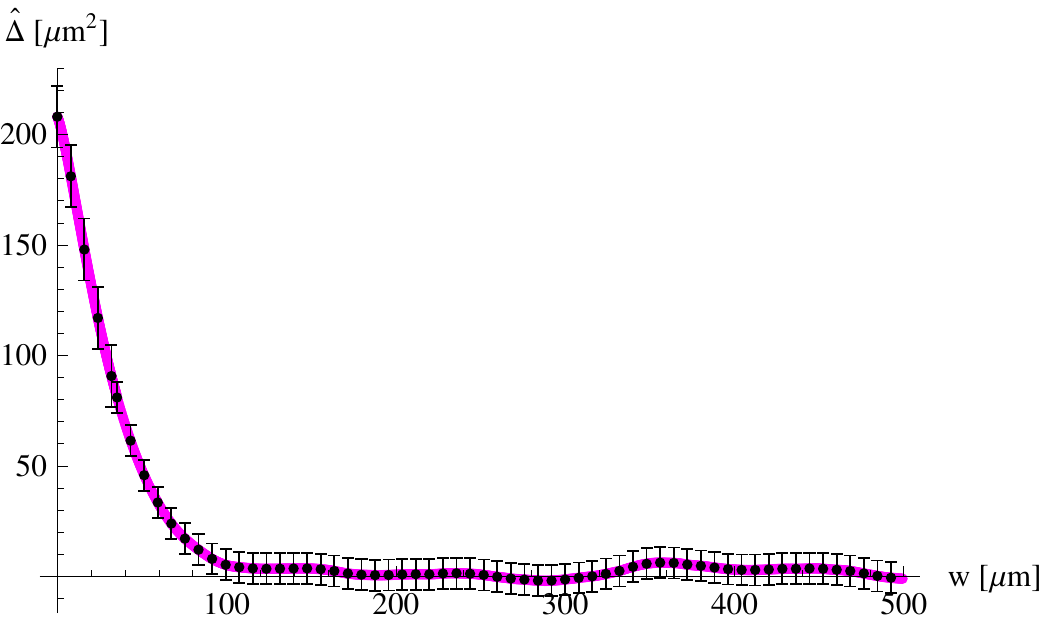}}}
\caption{Inset: The disorder correlator $\hat{\Delta} (w)$ for iso/Si
at $v=1\mathrm{\mu m/s}$ up to $w=35 \mathrm{\mu m}$, and then at $v=10 \mathrm{\mu m/s}$ for
$w>35 \mathrm{\mu m}$, with error-bars as estimated from the experiment. Main plot: The rescaled disorder correlator $\hat \Delta (w)/\hat \Delta (0)$ (green/solid) with error bars (red).  The dashed line is the 1-loop result from equation
(\ref{Delta1loop}).}
\label{Delta:iso/Si}
\end{figure}
We need to choose the CL length $l$ over which $h$ is averaged. It must be larger than the capillary length $L_c$, and as large as possible, while remaining  notably smaller than the plate size $L \approx 20$ mm, since strong distortions occur at the edge of the plate.
Moreover, slow changes of the plate properties cause slow variations of the capillary rise on the plate. These are small compared to the fluctuations of the \textit{local} CL height, but one expects that $\hat \Delta$ varies like $l^{-1}$ (for $l \gg L_c$). So, the larger $l$, the stronger the effect of the large-scale heterogeneities. As a compromise, we have chosen $l = 2L_C$. As shown in fig.~\ref{Delta(l)}, $\hat \Delta(0)$ has almost reached the asymptotic behavior for $l = 2L_c$.
Once $l$ is chosen, an accurate determination of $\hat \Delta(w'-w)$ requires that the CL explore a large number of pinned configurations. In other words,  the CL has to sweep a range of $w$ much larger than the width of the line. Following  \cite{MoulinetGuthmannRolley2002,MoulinetRossoKrauthRolley2004}, we define the CL width $W$ at a scale $l'$ as
$ W^2(l') = \langle(h(x,t)-h(x+l',t))^2 \rangle $, where the average is taken over $x$
and over all the successive configurations. For iso/Si, we find that the width $W \simeq 40\, \mathrm{\mu m}$  for $l'\gtrsim 2L_C$; for $\mathrm{H}_2$/Cs, the width is much smaller and difficult to measure, but we estimate that $W \approx 5\; \mathrm{\mu m}$ at large scale. The CL sweeps about 80 mm for iso/Si and about 5 mm for $\mathrm{H}_2$/Cs.
The plate displacement is thus much larger than $W$. This is necessary in order to get a reproducible shape for $\Delta$. As a consequence, the disorder and the substrate properties have to be fairly homogeneous over an area of at least $1\,\mathrm{cm}^2$, which is quite difficult to achieve. In particular, we have first tried to use water on glass decorated with Cr defects as in \cite{MoulinetGuthmannRolley2002}, but we could not get rid of strong variations of the capillary rise.

When comparing the two systems used in this work, the main difference lies in the defect size $\xi$, which is roughly one order of magnitude smaller for $\mathrm{H}_2$/Cs compared to iso/Si. The drawback of the low-temperature system is the poor characterization of the disorder which cannot be resolved optically. But the small value of $\xi$ is rather an advantage as the separation between the small-scale cut-off $\xi$ (defect size) and the large-scale  cut-off $L_c$ (capillary length) is about 3 orders of magnitude for $\mathrm{H}_2$/Cs, while only about 2 for iso/Si. As a consequence, it is difficult to observe the scaling regime for iso/Si. (For instance, $W(l')$ does not exhibit scaling in fig.~\ref{W(l)}). The case of $\mathrm{H}_2$/Cs is closer to previous experiments of the same type which did exhibit scaling.
%

We now discuss our experimental results and their comparison to theory. The raw data for the correlator $\hat \Delta(w)$ defined in (\ref{defdelta}) are shown in the insets of figs.~\ref {Delta:H2/Cs} and \ref{Delta:iso/Si}.
To compare with theory it is useful to define a dimensionless disorder correlator as in \cite{MiddletonLeDoussalWiese2006}
and plot the ratio $Y:=\hat \Delta(w)/\hat \Delta(0)$ as a function of the variable $z$ with $w=z \int_0^\infty \rmd w'\, \hat \Delta(w')/\hat \Delta(0)$ so that the area under the curve is unity. The resulting function has no more free parameter and
the functional RG theory of pinned systems predicts that in the limit of small $m$: (i) it should be the same function for all systems in a given universality class (UC); (ii)~this function should exhibit a linear cusp near $w=0^+$. In figs.~\ref {Delta:H2/Cs} and \ref{Delta:iso/Si}
 we have plotted these dimensionless correlators (solid lines).
\DRAFT{\tiny The scaling factor are... KAY: ??
Raw data are shown in the inset. The tail of $\hat \Delta(w)$
($w > 30\; \mathrm{\mu m}$) can be modeled as random
noise,allowing to estimate the statistical error **** Pierre: les erreurs ont sans doute ete surestimees
car d'apres Etienne elles sont proportionnelles a $\sqrt{w}$ ****
The plate velocity is
$V=3\; \mathrm{\mu m/s}$
The capillary number $\mathrm{Ca}:=\eta v/\gamma$ which is the ratio of viscous to capillary forces is about $2 \times 10^{-8}$ : the CL is close to the depinning transition and the function should be close to its fixed point. As predicted, we find a linear cusp in $\hat \Delta(w)$. }
Also shown in figs.~\ref {Delta:H2/Cs} and \ref{Delta:iso/Si}
 is the 1-loop prediction
\begin{equation}\label{Delta1loop}
z = \sqrt{Y_{\mbox{\scriptsize 1-loop}}-1-\ln Y_ {\mbox{\scriptsize 1-loop}}}\bigg/\int_0^1 \rmd y  \sqrt{y-1-\ln y}
\end{equation}
for the universality class described by Eq.~(\ref{a1}), i.e.~quasi-static depinning with irrelevant non-linear terms, the only class for which this function has been computed analytically yet. One sees  a very good agreement between data and this 1-loop FRG prediction. A similar agreement was observed numerically both in the statics and the dynamics of pinned systems with {\em local} elasticity \cite{MiddletonLeDoussalWiese2006,RossoLeDoussalWiese2006a}. At this stage we take this as a clear signature that we are dealing with a pinned system, and that a description using Eq.~(\ref{a1}) is possible.
A significantly higher precision and statistics (of a factor $\approx 10-20$) would be required to reach the one which was achieved in numerics \cite{MiddletonLeDoussalWiese2006,RossoLeDoussalWiese2006a}. That would provide a
decisive test on the universality class, and shed light on the debate about the large observed value for $\zeta \approx 0.5$ \cite{MoulinetRossoKrauthRolley2004}, while analytical predictions based on model (\ref{a1}) lie around $\zeta=0.4$ \cite{ChauveLeDoussalWiese2000a,RossoHartmannKrauth2002}. Closer examination of the data in fig.~\ref{Delta:iso/Si} shows that deviations from $Y_ {\mbox{\scriptsize 1-loop}}(z)$ can be mainly accounted for by rounding, which thus must be better controled and quantified.

\begin{figure}
\Fig{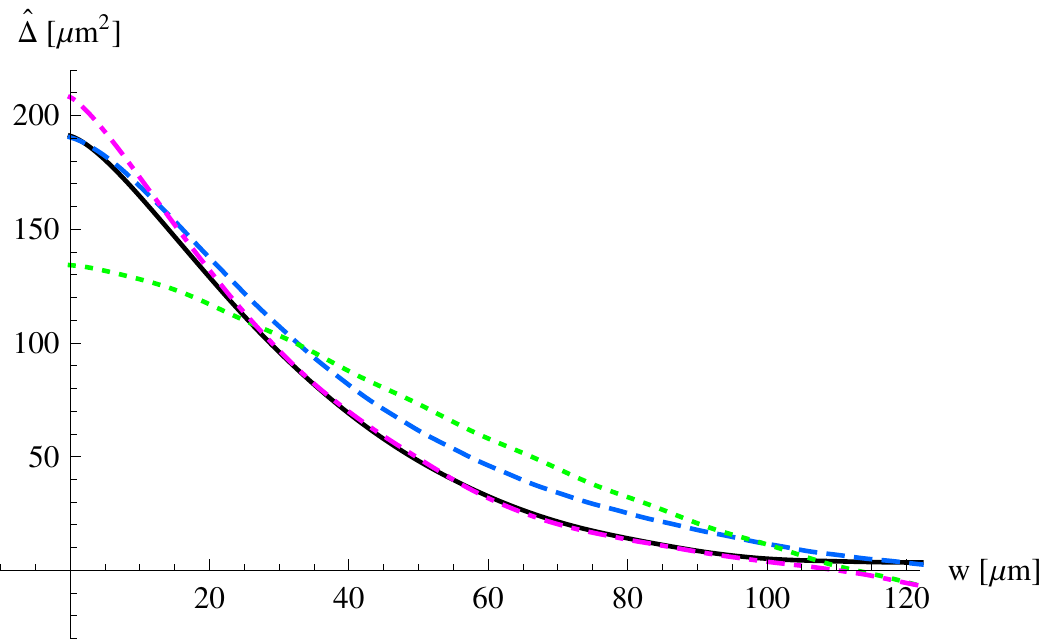}
\setlength{\inset}{5cm}
\Fig {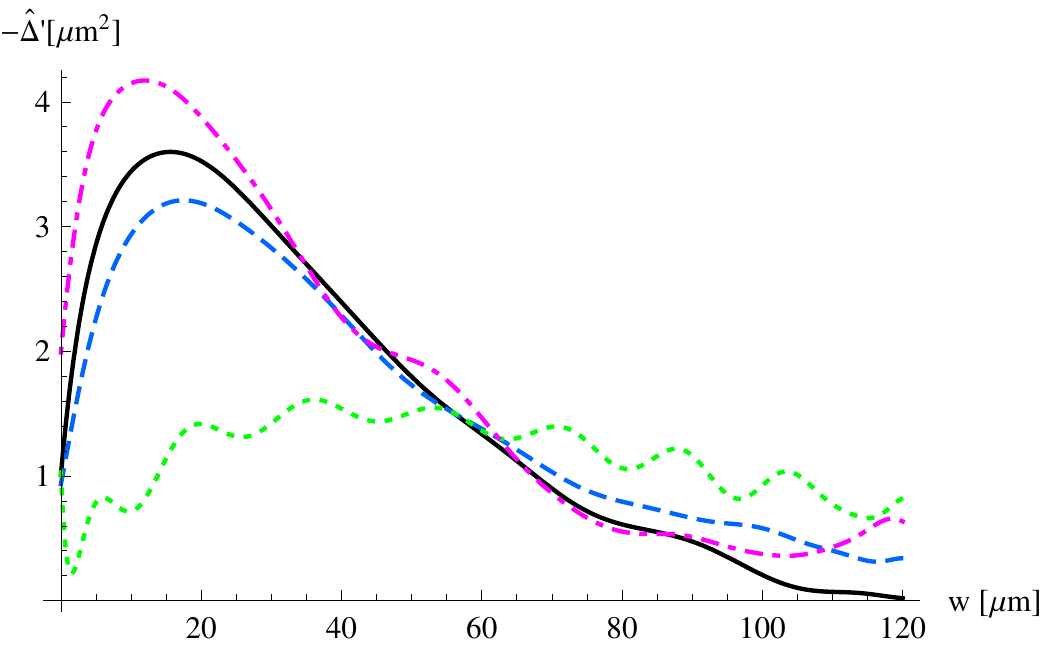}\hspace{-\inset}\raisebox{23mm}{\parbox[b]{\inset}{\fig{\inset}{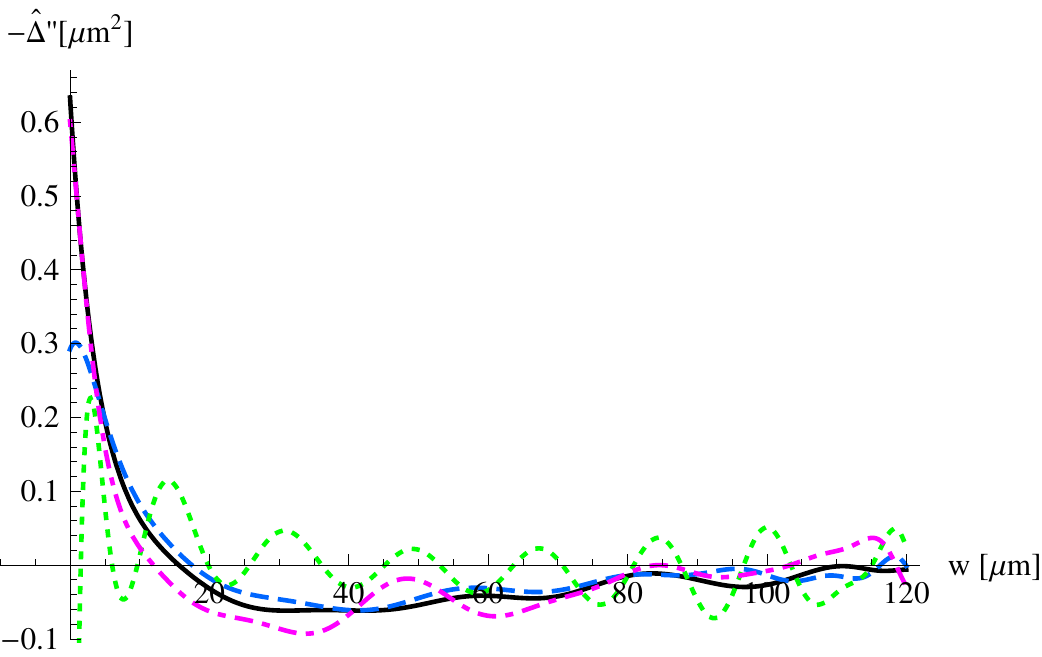}}}
\caption{The  disorder correlator $\hat \Delta (u)$ (top), and (minus) its first (bottom, main plot) and second (bottom inset) derivatives. The  velocities are $v=1\mathrm{\mu  m/s}$ (dashed-dotted/pink), $v=10\mathrm{\mu m/s}$ (solid/black), $v=100\mathrm{\mu
m/s}$ (blue/dashed) and
$v=1200 \mathrm{\mu m/s} $ (green/dotted). Statistical errors for $\Delta'(w)$ are $\approx 5\%$. (The derivatives are obtained
by fitting with a polynomial of degree 50 to 100 for $0\le w \le 120$.) The non-zero value
of $\hat\Delta'(0^+)$ is the signature of the predicted linear cusp.}
\label{f:diff-vels}

\label{f:diff-vels2}
\end{figure}

For iso/Si, we have performed the experiment at different velocities, in order to see the approach to the fixed point, and check for experimental artifacts.  This is shown in fig.~\ref{f:diff-vels}. We first comment on the shape of the correlator at the highest velocity ($v=1200\; \mathrm{\mu m/s}$). The flattening at the origin is expected as the system is driven away from the depinning threshold (i.e.\ away from the fixed point). However the large oscillations in $\hat \Delta(w)$ (better visible in its derivative) are surprising. We believe them to be due to surface waves of the liquid, possibly excited by the motor moving the plate.
Decreasing the velocity, the cusp at the origin becomes more and more pronounced, though even at the lowest velocity $v=1 \, \mu m$ the first derivative $\hat \Delta'(w)$ is not yet monotonically decreasing, as predicted by theory for $v=0^+$.

One possible origin of the observed rounding may be that the dissipation is not simply due to viscous shear in the meniscus but involves complex microscopic processes at the solid surface. This is demonstrated by the fact that the behavior of the correlator for vanishing $z$ is sensitive to the cleaning procedure of the plate, and by the fact that the homogeneous bare silicon surface displays an \textit{intrinsic} hysteresis (for a discussion, see \cite{MoulinetGuthmannRolley2004}). At this stage, we have little idea how the underlying microscopic disorder could change the behavior of the CL at the scale of the macroscopically patterned defects, and how to predict the possible resulting rounding effect.

On the other hand, there are also two fundamental reasons why the system could be slightly off the fixed point:
(i) rounding by the velocity, as observed above and expected form the theory (although most models predict that
$\Delta'(0^+)=0$ at $v>0$; this point is still debated \cite{chauve}) (ii) as suggested by the width data in fig.~\ref{W(l)}, the ratio  $\xi/L_C$ (or equivalently $m$) is yet too large to have reached the fixed point. The shape of $\hat \Delta'(w)$ is qualitatively what is expected: $\hat \Delta'(0^+)$ is strictly positive, hence the system is above the Larkin scale, but it is significantly smaller than its putative fixed-point value, since $\hat \Delta'(w)$  should be monotonically decaying there. Similarly $\hat \Delta''(w) <0$ near $w=0$ while the fixed-point value is expected to be positive.
This interpretation, if confirmed, is interesting, as it implies that $\hat \Delta(w)$ is a sensitive new probe, which can be made quantitative, to test how far the system is from criticality. 

\begin{figure}
\setlength{\unitlength}{1mm}
{\begin{picture}(85,50)
\put(0,0){\fig{7cm}{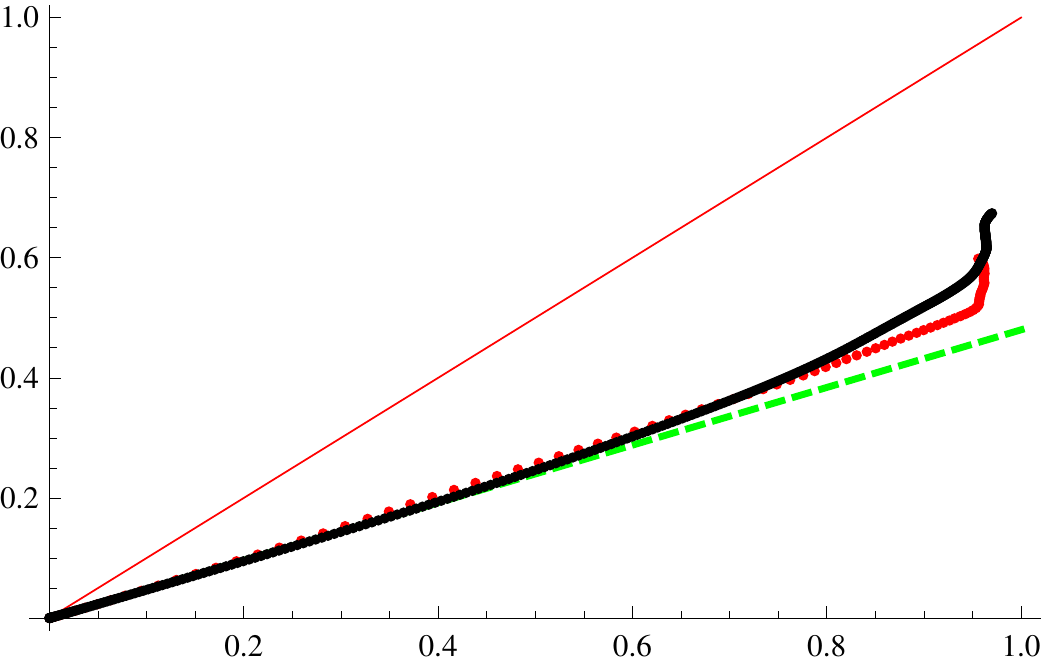}}
\put(70,2){$\left[1-\frac{\hat \Delta(w)}{\hat \Delta(0)} \right]^2 $}
\put(0,47){$\frac{\hat Q(w)}{\hat \Delta(0)^2}$}
\end{picture}}
\caption{Dimensionless parametric plot of $\hat Q (w)/\hat \Delta (0)^{2}$ versus
$\left[1-\frac{\hat \Delta (w)}{\hat \Delta (0)} \right]^{2}$ for both
experiments $\mathrm{H}_2$/Cs (red/grey) and iso/Si at $v=10\mathrm{\mu m /s}$ (black). The thin red
(grey) line is the mean-field prediction. The thick dashed green (grey) line is an extrapolation
based on a loop expansion, as discussed in the text.}
\label{3cum}
\end{figure}

More information can be obtained from the experimental data by computing the third cumulant:
\begin{equation}\label{}
\hat{S}_{3} (w-w') : =\left<  \left[{\sf h}_l(w)-{\sf h}_l(w') \right]^{3}\right>
\end{equation}
It is convenient to plot the function
\begin{equation}\label{13}
\hat Q (w):= \frac{1}{6} \int_{0}^{w} \rmd w' \hat{S}_{3} (w')\ ,
\end{equation}
Indeed, the FRG predicts that
\begin{equation}\label{14}
\frac{\hat Q (w)}{\hat \Delta(0)^2} \approx A \left[1-\frac{\hat \Delta (w)}{\hat \Delta(0)} \right]^{2}\ ,
\end{equation}
with an exact equality and the universal amplitude $A=1$
at the mean-field level (i.e.\ resummation of tree diagrams, $d \geq d_{\mathrm{uc}} = 2$).
The proportionality (with no attempt to measure $A$) was checked
numerically \cite{RossoLeDoussalWiese2006a} for depinning with {\em local} elasticity.
In fig.~\ref{3cum} we plot both sides of eq.~(\ref{14}): we see that the proportionality holds very nicely and that the two experiments fall on top of each other, confirming the universality of the
slope $A$, which is measured to be $A\approx 0.53$.  The FRG calculation, using the scaled elastic kernel $\tilde \epsilon_{\pi/2}(x)$ yields $A=1/(1+ \frac{8}{9} \epsilon)+O(\epsilon^2)$ with $\epsilon=2-d$ (see (E12) of \cite{LeDoussalWiese2008c}) which yields $9/17=0.53$ and $0.11$ for the two Pade approximants at $\epsilon=1$, while
the kernel $\tilde \epsilon_{\theta}(x)$ with $\theta=40^\circ$ 
gives \cite{LeDoussalWieseToBePublished} $A=0.48 \pm 0.13$ (shown in fig.\ \ref{3cum}). These values are reasonable, given that the correction
to mean field is large and one cannot hope for high precision.
\DRAFT{\footnote{**** for info, not a final version **** the predictions are $A=1 - \frac{2}{9} \epsilon$ (local),
$A=1 - \frac{4}{9} \epsilon$ ($|q|+1$ elasticity) $Pade(1,0)=0.56$, $Pade(0,1)=0.69$
$A=1 - \frac{8}{9} \epsilon$ ($\sqrt{q^2+1}$ elasticity)  $Pade(1,0)=1/9=0.11$, $Pade(0,1)=9/17=0.53$}}
Also note that deviations from the functional form  in Eq.~(\ref{14})  are
expected at 1-loop order, but they should be small as found numerically in \cite{RossoLeDoussalWiese2006a}. We conclude that the  agreement between experiment and  theory is satisfactory for the third cumulant.
\DRAFT{
\footnote{****Pierre: to answer Etienne (i) simple proportionality in (\ref{14}) is
a check of a functional form, very much as what is done for the rescaled $Y=\Delta/\Delta(0)$ versus $z$ function. Predicting $A$ is more difficult as it is an amplitude (ii) le resultat pour $Y$ est "one loop" dans le sens qu'il demande de resoudre le point fixe de $\Delta$ au premier ordre en $\epsilon$, donc il donne le resultat exact pour $d=d_{uc}=2$ alors que pour $d>2$ on aurait une fonction non universelle. Par contre on peut montrer que la formule (\ref{14}) reste vraie a $m$ petit pour $d \geq 4$ meme si chaque cote est non universel, d'ou la denomination mean-field}}


The properties of the CL can also be characterized by the distribution of
the sizes $S$ of avalanches, or forward jumps (see fig.~\ref{h(t)}), where by definition
$S$ is the area swept during the avalanche. At the
critical point, $m= 0$, the distribution is expected to be a power law
characterized by an exponent $\tau$. At small $m>0$, the correlation
length is finite and the avalanche-size distribution $P(S)$ for $S \gg
S_{\mathrm{min}}$  is cut off at scale
\begin{equation}\label{4}
S_m: =\frac{\langle S^2 \rangle}{ 2 \langle S\rangle} \ ,
\end{equation}
\begin{figure}[t]
\Fig{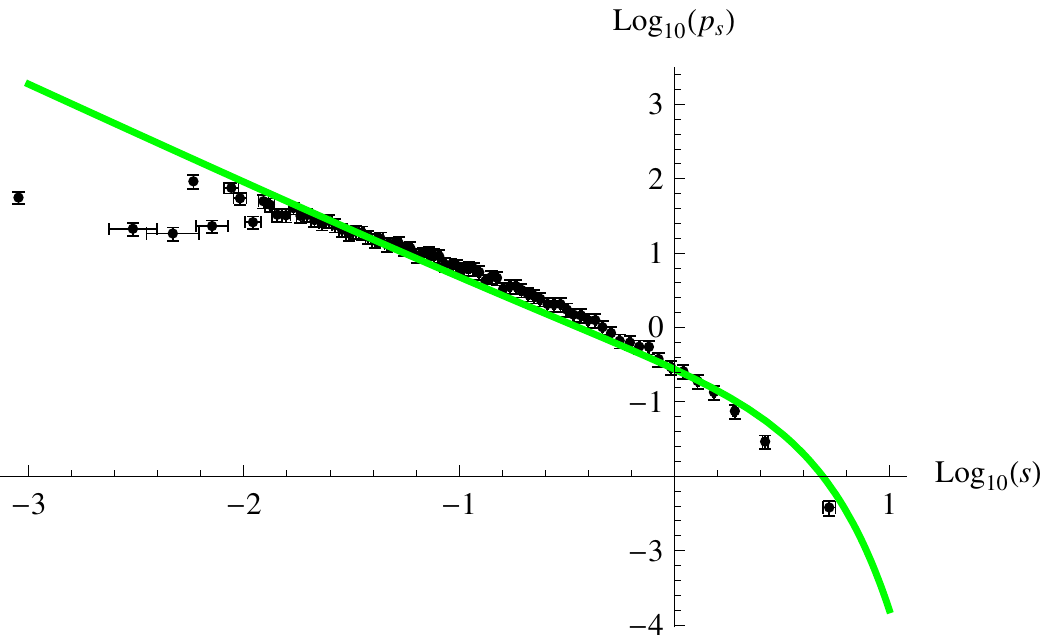}
\caption{The measured dimensionless avalanche-size distribution function $p(s)$, as defined in the text, for iso/Si. For comparison an analytical estimate, defined
in Eq.~(\ref{avals}), based on a one loop calculation with the scaled elastic kernel $\tilde \epsilon_{\pi/2}(x)$.
The small-size cutoff is $\log_{10} ( s)=-2$, corresponding to the size of a defect.}
\label{f:ps}
\end{figure}%
where here and below $\langle \ldots \rangle$ denotes an average over $P(S)$.
One expects that in the variable $s:=S/S_m$ the avalanche-size distribution
exhibits universality, i.e.\ independence of short scales, and that, for $1<\tau<2$, it takes the form
\begin{equation}\label{distrib}
P (S) \rmd S := \frac{\left< S \right>}{S_m} \, p\! \left(\frac{S}{S_{m}}\right)  \frac{\rmd S}{S_m}\ .
\end{equation}
The function
$p(s)$
is universal, and by construction from (\ref{4}) and (\ref{distrib}) normalized s.t.\
$\int_0^\infty \rmd s\, s p(s) =1$ and $\int_0^\infty \rmd s\, s^2 p(s) =2$ \cite{ps}. The analytical prediction \cite{LeDoussalWiese2008c}
for $p(s)$ reads, based on the model (\ref{a1}) and with the scaled elastic kernel $\tilde \epsilon_{\pi/2}(x)$:
\begin{equation}\label{avals}
p(s) = A'   s^{-\tau  } \exp\!\bigg ( -\frac{B'}{4}
   s^{\delta '}+ C' \sqrt{s} - D' s^{\frac32}\bigg)
\end{equation}
with $B' =1+ \frac{1}{3} (\gamma -2) \epsilon$, $C'=\frac{\sqrt{\pi
 } \epsilon }{3}$, $D'=  \frac{\sqrt{\pi } \epsilon }{36}$, $A'$
 given by the normalization and the exponent $\delta'= 1 + \frac{\epsilon}3$.
 This prediction is exact to first order in $\epsilon=2-d$, and to produce our
 analytical estimate we set $\epsilon=1$  and rescale both
 axis to ensure the two normalization conditions. For the $\tau$ exponent, we
 have used the conjectured relation $\tau(\zeta) =2-1/ (d+\zeta)$ since we proved \cite{LeDoussalWiese2008c}
 that the latter is exact at least to one loop. Inserting the measured $\zeta \approx 0.5$
 yields the prediction $\tau \approx 4/3$. 
  \DRAFT{Kay: on peut fitter avec tout, donc ca n'a pas trop de sens. j'ai teste zeta=0.5 et zeta=0.33.
**** Pierre: (i) commenter si ce $\tau$ fitte la manip et barres d'erreur (ii) si c'est le $\zeta$
 d'une autre manip c'est inconsistent, peut on plutot estimer le $\zeta$ effectif pour cette manip  ???
 ETIENNE : je suis d'accord. MAIS mettre cette discussion après les considérations expérimentales
 (iii) Kay peux tu essayer $\zeta=0.4$ et voir la difference. Dans chaque cas
 merci de donner les valeurs numeriques de $A',B',C',D'$ que je puisse verifier merci ****}

\begin{figure}
\Fig{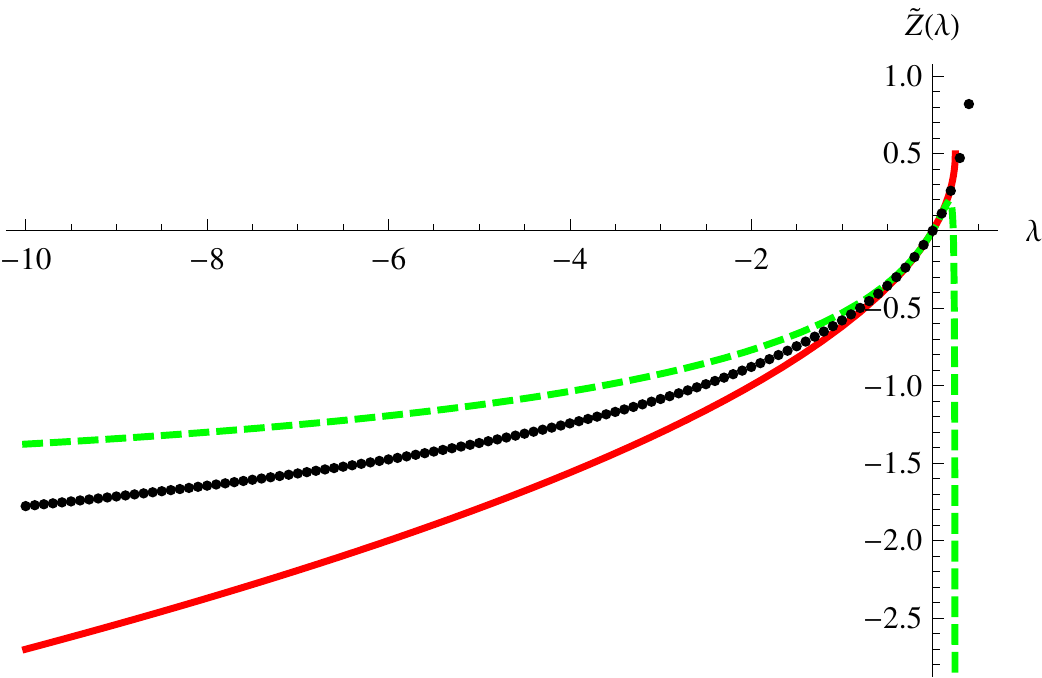}
\caption{$\tilde Z (\lambda)$ defined in (\ref{a8}), from bottom to top: the mean-field prediction (red, solid), the experimental data (black dots), and the extrapolation based on the one loop calculation for a scaled elastic kernel $\tilde \epsilon_\theta(x)$ with $\theta=40^\circ$  (green, dashed).}
\label{f:tZlambda}
\end{figure}

The size distribution is measured only for iso/Si with a standard camera
(acquisition rate: 15 Hz),
at a velocity $v=1\;\mathrm{\mu m/s}$. We consider that an event occurs if the local displacement between
two successive images is larger than a threshold $\delta h  \simeq 8\, \mathrm{\mu m}$. This threshold is about the defect size,
which is the smallest displacement occuring in an avalanche, and is about 10 times larger
than the resolution on the CL position.
The area $S$ swept in a single avalanche is then computed as the sum of the displacements occuring before the CL is pinned again. We detected $16 \times 10^3$ avalanches for a swept area of $3.5 \times 20 \;\mathrm{mm}^2$.
We have checked that the resulting distribution is not sensitive to the value of the threshold and that the acquisition rate is fast enough to avoid lumping together uncorrelated avalanches.

Experimental data and predictions are shown in fig.~\ref{f:ps}. We see that the parameter-free scaling
collapse is quite good down to $s=10^{-2}$. This value of $s$
corresponds  to $S=100\, \mathrm{\mu m}^{2}$, which is the cutoff
 expected from the size of the defects whose area is $100\, \mathrm{\mu m}^{2}$.
 However our accuracy is limited by the available sizes, and e.g.\  not sufficient to discriminate between $\tau(\zeta=0.5)=4/3$, and $\tau(\zeta=0.4)=1.29$.
 \DRAFT{
 ******Etienne : je suggère de laisser tomber la discussion oiseuse sur le sous echantillonnage
 qui n'a plus guere d'intéret depuis que j'ai refait des manips à basse vitesse.
 Par ailleurs, c'est ICI qu'il faut discuter dans quelle mesure p(s) est sensible à la valeur de $\tau$ (très peu
 autant que je me souvienne)******
 }
While $p(s)$ has strong statistical fluctuations, the characteristic function
\begin{equation}\label{a8}
\tilde Z (\lambda) := \int_{0}^{\infty} \rmd s \, p (s)\left(\rme^{\lambda s}-1 \right)
\end{equation}
can be measured quite accurately as shown in fig.~\ref{f:tZlambda}. The 1-loop extrapolation ($\epsilon=1$)
is closer to the data than the mean-field result and using the elasticity $\tilde \epsilon_{\theta=40 ^\circ}(x)$
(shown in fig.~\ref{f:tZlambda}) is also closer than using $\tilde \epsilon_{\theta=\pi/2}(x)$ (not shown). Hence the
comparison to field-theory predictions \cite{LeDoussalWiese2008c,LeDoussalWieseToBePublished} is satisfactory.

We now come to a final test of the avalanche picture underlying the FRG calculations.
According to \cite{LeDoussalMiddletonWiese2008,LeDoussalWiese2008c}, the slope $\hat \Delta'(0^+)$ of the linear cusp  is proportional to the scale $S_m$ of the avalanche-size distribution, 
\begin{equation}\label{}
L \left|\hat \Delta' (0^{+}) \right| =S_{m}\equiv  \frac{\left< S^{2} \right>}{2\left< S \right>}\ . 
\end{equation}
We find $S_{m} \simeq 9000 \mathrm{\mu m}^{2}$.
At the lowest velocity,  $|\hat \Delta'(0^+)|=1.75 \mathrm{\mu m}^2$ and $L = 3500 \mathrm{\mu m}$, one finds $L|\hat \Delta' (0^{+}) |\simeq 6100\,\mathrm{\mu m}^{2}$. However, estimating $\hat \Delta' (0^{+})$ is difficult because we do not understand the origin for the rounding of $\hat \Delta(w)$ when $w \rightarrow 0$. An upper bound for
$|\hat \Delta' (0^{+}) |$ is the slope of $\hat \Delta'$ at the inflection point, which yields $L|\hat \Delta' (0^{+}) | < 14000\,  \mathrm{\mu m}^{2}$.
The third moment of the avalanche-size distribution can similarly be related to the third cumulant of the center-of-mass fluctuations:
$\left< S^3 \right> \left< S\right>/(3 \left< S^2 \right>^2)=A$, with $A$ defined in eq.~(\ref{14}). From $p(s)$ we find $A = 0.77$, while the relation between $\hat Q(w)$ and $\hat  \Delta(w)$ yields $A=0.53$.
The agreement in both cases is only fair and more experiments are needed. In particular, a smaller ratio $\xi/L_C$ would be helpful.

To conclude: By examining the fluctuations of the mean height of the contact-line at depinning, we have measured the renormalized disorder-correlator $\Delta(w)$. The latter is the central object of the functional RG field theory, and its predicted cusp, which is the sign of metastability, shocks and avalanches, was under intense debate from the field theory side. Here we have made the first
 comparison between experiment and theory. It shows qualitatively and quantitatively that the ideas in the latter are correct, and opens new ways of quantifying the former, calling for new experiments. 

We thank G.\ Borot and M.\ Pettersen for help with the experiments and A.\ Rosso for discussions.

\end{document}